\documentclass[12pt]{article}
\usepackage{epsfig,cite,amsmath,color}

\input paperdef
\evensidemargin \oddsidemargin
\makeatletter
\renewcommand{\section}{\@startsection {section}{1}{\z@}%
                           {-3.5ex \@plus -1ex \@minus -.2ex}%
                           {2.3ex \@plus.2ex}%
                           {\mathversion{bold}\normalfont\Large\bfseries}}
\renewcommand{\subsection}{\@startsection{subsection}{2}{\z@}%
                           {-3.25ex\@plus -1ex \@minus -.2ex}%
                           {1.5ex \@plus .2ex}%
                           {\mathversion{bold}\normalfont\large\bfseries}}
\renewcommand{\subsubsection}{\@startsection{subsubsection}{3}{\z@}%
                           {-3.25ex\@plus -1ex \@minus -.2ex}%
                           {1.5ex \@plus .2ex}%
                           {\mathversion{bold}\normalfont\normalsize\bfseries}}
\makeatother

\allowdisplaybreaks

\begin{document}
\thispagestyle{empty}

\def\thefootnote{\fnsymbol{footnote}}

\begin{center}

{\large\sc 
\bf{Exclusive MSSM Higgs production at the LHC after Run I}}

\vspace{1cm}

{\sc 
M.~Tasevsky%$^{4}$%
\footnote{email: Marek.Tasevsky@cern.ch}%
}

\vspace*{0.5cm}

{\sl
Institute of Physics of the Academy of Sciences of the Czech Republic, 
18221 Prague 8, Czech Republic

}

\end{center}

\begin{abstract}
We investigate the prospects for Central Exclusive Production (CEP)
of MSSM Higgs bosons at the LHC using forward proton detectors 
proposed to be installed at 220~m and 420~m distance around ATLAS and / or CMS.
We summarize the situation after the first and very successful data taking 
period of the LHC. The discovery of a Higgs boson and results from searches
for additional MSSM Higgs bosons from both the ATLAS and CMS experiments, based 
on data samples each corresponding to about 25~fb$^{-1}$, have recently led
to a proposal of new low-energy MSSM benchmark scenarios. The CEP signal cross 
section for the process $H/h \to b \bar b$ and its backgrounds are estimated in
these new scenarios. We also make some comments about the experimental 
procedure if the proposed forward proton detectors are to be used to 
measure the CEP signal. 

\end{abstract} %end of abstract

\def\thefootnote{\arabic{footnote}}
\setcounter{page}{0}
\setcounter{footnote}{0}

\newpage

\section{Introduction}
\label{intro}
The interest in the central exclusive production of new particles is still
significant over the last decade \cite{KMR0,ar,KMRProsp,acf,DKMOR,KMRbsm,HarlandLang:2013jf}. 
The process is defined as $pp\rightarrow p\oplus\Phi\oplus p$
where all of the energy lost by the protons during the interaction
(a few per cent) goes into the production of the central system, $\Phi$. The 
final state therefore consists of a centrally produced system (e.g. dijet, 
heavy particle or Higgs boson) coming from a hard subprocess, two very forward 
protons and no other activity. The '$\oplus$' sign denotes the regions devoid 
of activity, often called rapidity gaps. A simultaneous detection of both 
forward protons and the central system opens up a window to a rich physics 
program covering not only exclusive but also a variety of QCD, Electroweak and
beyond Standard Model (BSM) processes (see e.g. 
\cite{KMRProsp,CMS-Totem,FP420,AFP,PPS,cr,kp1,bcfp,mt1}). Such measurements can
put constraints on the Higgs sector of Minimal Supersymmetric SM (MSSM) and 
other popular BSM scenarios \cite{KKMRext,diffH,diffH2,CLP,fghpp,ismd,tripl,eds09,dis11,je2,CEPW,Djouadi}. In SM the CEP of Higgs boson has been studied in 
\cite{HarlandLang:2013jf,CMS-Totem,cr1,ATLASnote,smhww,ISMD05}. 

CEP is especially attractive for three reasons: firstly, if the 
outgoing protons remain intact and scatter through small angles then, to a very 
good approximation, the primary di-gluon system obeys a $J_z=0$, $\cC$-even, 
$\cP$-even selection rule~\cite{KMRmm,KKMR}. Here $J_z$ is the projection of 
the total angular momentum along the proton beam axis. This allows therefore 
a clean determination of the quantum numbers of any observed resonance. Thus, 
in principle, only a few such events are necessary to determine the quantum 
numbers, since the mere observation of the process establishes that the 
exchanged object is in the  $0^{++}$ state. Secondly, from
precise measurements of the proton momentum losses, $\xi_1$ and
$\xi_2$, and from the fact that the process is exclusive, the mass
of the central system can be measured much more precisely than from
the dijet mass measured in the central detector, by the so-called
missing mass method, $M^2=\xi_1\xi_2 s$ ($s$ is the square of the proton-proton 
center-of-mass energy) which is independent of the
decay mode. Thirdly, in CEP, in particularly so far elusive $b\bar b$ mode, the 
signal-to-background (S/B) ratios turn out
to be close to unity, if the contribution from pile-up is not considered. 
This advantageous signal-to-background ratio is due to the
combination of the $J_z=0$ selection rule, the potentially excellent
mass resolution, and the simplicity of the event signature in the
central detector. Another important feature of forward proton
tagging is the fact that it enables the strongest decay modes, namely $b\bar b$,
$\WW$ and $\tau\tau$ to be observed in one process. In this way, it may be 
possible to access the Higgs boson coupling to bottom quarks. This may be 
challenging in conventional search channels at LHC due to large QCD 
backgrounds although $\hbb$ is the dominant decay mode for a light 
SM Higgs boson. Here it should be kept in
mind that access to the bottom Yukawa coupling will be crucial as an
input also for the determination of 
Higgs couplings to other particles~\cite{HcoupLHCSM,HcoupLHC120,LHCHiggsX1,LHCHiggsX2}.
As already discussed in \cite{diffH2}, measuring the proton transverse 
momentum and azimuthal angle $\phi$ distributions in forward proton detectors 
would enable us to search for a possible $\cp$-violating signal in the Higgs
sector \cite{kmrcp}. As shown in \cite{kmrcp} the contribution caused by the
$\cp$-odd term in the $gg \rightarrow H$ vertex is proportional to the 
triple-product correlation between the beam direction and the momenta of 
outgoing detected protons. In some $\cp$-violating MSSM scenarios 
\cite{je2,CEPW} an integrated counting asymmetry (based on counting events with
$\phi > \pi$ and with $\phi < \pi$) can be sizable. 

Studies of the Higgs boson produced in CEP used to form a core of the physics 
motivation for upgrade projects to install forward proton detectors at 420~m 
\cite{FP420} and 220~m from the ATLAS (AFP project \cite{AFP}) and CMS 
(PPS project \cite{PPS}) detectors. At the moment, only 220~m stations 
are considered to be installed in ATLAS and CMS. 

As it is well known, many models of new physics require an extended
Higgs sector. The most popular extension of the SM is the 
MSSM~\cite{susy1,susy2,susy3}, 
where the Higgs sector consists of five physical states (two Higgs doublets are
required). At the lowest order the 
MSSM Higgs sector is $\cp$-conserving, containing two $\cp$-even bosons, the 
lighter $h$ and the heavier $H$, a $\cp$-odd boson, $A$, and the charged bosons
$H^\pm$. It can be specified in terms of the gauge couplings, the ratio of the 
two vacuum expectation values, $\tb \equiv v_2/v_1$, and the mass of the $A$
boson, $\MA$. The Higgs sector of the MSSM is affected by large higher-order
corrections (see for example \cite{reviews1,reviews2,reviews3,reviews4} for 
reviews), which have to be taken into account for reliable phenomenological 
predictions. 

A brief summary of what has been studied and published in previous texts
regarding the prospects of the CEP processes is given below.  
In \cite{diffH} and \cite{diffH2} the CEP of the Higgs boson in MSSM
at the LHC energies was studied in great detail. 
In \cite{diffH} the detailed description is given of how the signal and all 
relevant background processes to all three decay modes are calculated. We 
plotted the enhancement factors of MSSM/SM cross sections and statistical 
significances in the ($\MA$, $\tb$) planes together with the LEP 
exclusion regions for the small $\alpha_{\rm eff}$ scenario in the case of the 
WW decay
and for the $\Mhmax$ and nomixing scenarios, defined in \cite{benchmark2}, in 
the case of the $b\bar b$ and $\tau\tau$ decays. In \cite{diffH2} the results 
in the $\Mhmax$ and nomixing scenarios were updated following the development 
on both the theory and experimental procedure side. On the theory side we have 
taken more accurate calculations of the process associated with bottom-mass
terms in the Born amplitude contributing to the total background of the 
$b\bar b$ mode \cite{shuv,screen} and used an improved version of the code 
{\tt FeynHiggs}~\cite{feynhiggs1,feynhiggs2,mhiggslong,mhiggsAEC,mhcMSSMlong} 
employed for the cross section and
decay width calculations. The development on the experimental side was 
reflected by adding the Tevatron MSSM exclusion regions corresponding to 
Tevatron searches for MSSM Higgs bosons. The exclusion regions were evaluated  
with {\tt HiggsBounds}~\cite{higgsbounds1,higgsbounds2}. Besides the $\Mhmax$ 
and nomixing 
scenarios other scenarios have been investigated, namely those yielding the
correct amount of the cold dark matter abundance, the so called Cold Dark 
Matter (CDM) scenarios (see \cite{EWPOBPOCDM,CDM} for more details), and in 
addition in one model beyond SM, the so called SM4 model (see e.g. 
\cite{four-gen-and-Higgs}) with a fourth generation of quarks and leptons. 
While the SM4 model is practically ruled out \cite{Doujadi-Lenz} by the recent LHC 
measurements of Higgs-mediated cross sections \cite{SM4-indirect1,SM4-indirect2}
and direct searches \cite{SM4-direct1,SM4-direct2}, the CDM scenarios are still 
viable. In \cite{diffH2} we also stressed the importance and advantages of the 
CEP process in the determination of spin-parity quantum numbers of Higgs 
bosons. 

The motivation and organization of this text is the following. Last year, the 
discovery of a new resonance with mass close to 125.5~GeV has been announced by 
the ATLAS \cite{HdiscA} and CMS \cite{HdiscC} experiments. Preliminary estimates
of the spin-parity (by ATLAS \cite{JATLAS}, CMS \cite{JCMS} and Tevatron 
\cite{TevHiggsSM}) of this resonance and its couplings (by ATLAS 
\cite{CouplATLAS} and CMS \cite{CouplCMS}) suggest that it is a Higgs boson 
with properties similar to the SM Higgs boson. Proving, however, that it is the 
Higgs boson coming exclusively from the SM, MSSM or other BSM theories will 
require measuring precisely its spin, $\cp$ properties, mass, width and 
couplings which is a program for the next decade or so. 
%(the Higgs signal strength close to unity, Higgs couplings and spin/parity 
%consistent with those in SM) \cite{LHC-SMHiggs}. 
At the same time,
results from numerous analyses searching for the MSSM signal at LHC have been 
published and further analysed. Based on these results i) new low-energy MSSM 
benchmark scenarios have been proposed \cite{newscenarios,LHCHiggsX2} that are 
compatible over large parts of the ($\MA$, $\tb$) parameter plane 
with the mass and production rates of the observed Higgs boson signal at 
125.5~GeV, and ii) the most recent LHC exclusion regions have been evaluated 
using the latest version of the program {\tt HiggsBounds} 
\cite{higgsbounds1,higgsbounds2}. 
The aim of this analysis is to
investigate the CEP of Higgs boson in these new benchmark scenarios 
taking into account the recent LHC exclusion regions and the region of the 
allowed Higgs mass.  

\subsection*{Luminosity scenarios}\label{lumiscenarios}
Where relevant, i.e. in benchmark scenarios where the statistical significances 
and signal event yields are sufficiently large, we consider two scenarios for
integrated luminosity and experimental conditions for CEP processes at the LHC.
As explained for example in \cite{diffH,ismd,eds09}, one of the two 
luminosity scenarios is to show a physics gain that could be expected if event
rates are higher by a factor of~2 (due to improvements on the experimental 
procedure side and possibly higher signal rates), denoted by ``eff$\times 2$''.
We furthermore assume for sake of simplicity a center-of-mass energy of
$\sqrt{s} = 14 \tev$ at the LHC. Lower energies and correspondingly
lower cross sections would require correspondingly higher luminosities.

\newcommand{\sixooo}{600 \ifb}
\newcommand{\sixoooeff}{600 \ifb\,eff$\times2$}

\begin{itemize}
%\item \underline{\sixoo :}\\[.4em] 
%An integrated LHC luminosity of ${\cal L} = 2 \times 30~{\rm fb}^{-1}$,
%corresponding roughly to three years of running at an instantaneous
%luminosity
%${\cal L} \sim 10^{33} \, {\rm cm}^{-2} \, {\rm s}^{-1}$ by both ATLAS
%and CMS. With such a luminosity the effect of pile-up is not negligible
%but can be safely kept under control. 

%\medskip
%\item \underline{\sixooeff :}\\[.4em] 
%The same integrated LHC luminosity as in the above scenario but with 
%event rates that are higher by a factor of 2 (see the discussion of possible
%improvements and theoretical uncertainties in \cite{diffH}). 

\medskip
\item \underline{\sixooo :}\\[.4em] 
An integrated LHC luminosity of ${\cal L} = 2 \times 300~{\rm fb}^{-1}$,
%and the same efficiency factors as in the scenario with
%${\cal L} = 60~{\rm fb}^{-1}$.
corresponding roughly to three years of running at an instantaneous
luminosity \\
$\cL \approx 10^{34}\,{\rm cm}^{-2} \, {\rm s}^{-1}$ by both ATLAS and CMS.

\medskip
\item \underline{\sixoooeff :}\\[.4em] 
The same integrated LHC luminosity as in the scenario with 
${\cal L} = 2 \times 300~{\rm fb}^{-1}$ but with event rates that are higher by 
a factor of 2 (see the discussion of possible improvements and theoretical 
uncertainties in \cite{diffH}).
\end{itemize}

%%%%%%%%%%%%%%%%%%%%%%%%%%%%%%%%%%%%%%%%%%%%%%%%%%%%%%%%%%%%%%%%%%%%%%%%%%%%%%%
%%%%%%%%%%%%%%%%%%%%%%%%%%%%%%%%%%%%%%%%%%%%%%%%%%%%%%%%%%%%%%%%%%%%%%%%%%%%%%%

\section{Update of the MSSM analysis}
\label{sec:update}

A detailed description of the procedure to obtain predictions of the signal
and background cross sections for all three Higgs boson decay modes is given
in \cite{diffH}. Their update can be found in \cite{diffH2}. The results 
shown here differ from \cite{diffH,diffH2} mainly in the choice of MSSM 
benchmark scenarios and by adding the LHC exclusion regions and the allowed 
Higgs mass region. The formulas to calculate the signal and backgrounds in SM
are the same as used in \cite{diffH2} (see the discussion below). Predictions 
within {\tt FeynHiggs} have been updated (from the version 2.7.1 used in 
\cite{diffH2} to the version 2.9.4 used here) but the changes in 
{\tt FeynHiggs} had practically no impact on our analysis. 
We have as well updated the {\tt HiggsBounds} program (from the version 1.2.0 
used in \cite{diffH2} to the version 4.0.0 used here) which is used to evaluate
the LHC MSSM exclusion regions - they supersede the Tevatron exclusion regions
used in \cite{diffH2}.

%%%%%%%%%%%%%%%%%%%%%%%%%%%%%%%%%%%%%%%%%%%%%%%%%%%%%%%%%%%%%%%%%%%%%%%%%%%%%%%

\subsection*{Calculation of signal and \boldmath{$b\bar b$} background}

The signal cross section is calculated on the basis of the 
prediction for the CEP of a SM Higgs boson together with an
appropriate rescaling using the partial widths of the neutral $\cp$-even
Higgs bosons of the MSSM into gluons, $\Ga(\phi \to gg)$ ($\phi = h,H$),
as implemented in {\tt FeynHiggs} (details are given in \cite{diffH}).

Since the publication of \cite{diffH} there was a fair development in the 
calculations of the CEP cross sections concerning both, the hard matrix element
(see \cite{acf,mrw,cf,KMRIns,KRS,Pasechnik}) and the so-called soft absorptive 
corrections and soft-hard factorization breaking effects (see \cite{kmrf,nns2} 
for details and references). As discussed in \cite{HarlandLang:2013jf} recent 
calculations of the combined enhanced and eikonal soft survival factor give 
lower values than the value 0.03 we used in \cite{diffH}. Also taking a more
appropriate factorization scale $M$ (rather than $\approx 0.62 M$) in 
calculating Sudakov suppression almost halves the CEP cross section (of both,
the signal and background) \cite{cf}. 
On the other hand as discussed in \cite{KMRH,HarlandLang:2010ep} we may expect
the cross section to be increased by higher order corrections and by using the
CTEQ6L \cite{CTEQ6L} LO proton PDF that give the best agreement of the CEP
calculations with CDF data on the exclusive $\gamma\gamma$ production 
\cite{CDFgg}. A combined effect of all changes is estimated to be rather small.
With the current theory accuracy of a factor of $\sim 2.5$, we therefore do not
feel the necessity to revise the procedure to calculate the exclusive 
$gg$-luminosity, which determines the rates of signal and background events
as used in \cite{diffH}. Ways to test the theoretical formalism at LHC, 
with or without forward proton tagging, are summarised in \cite{early}.

As explained in \cite{diffH2} we use the improved background formula where the
NLO corrections to the bottom-mass terms in the Born amplitude were added
following the results of \cite{shuv}. For the term associated with the
prolific $gg^{PP}\to gg$ subprocess which can mimic $b\bar b$ production due
to the misidentification of the gluons as $b$ jets, we still use $P_{g/b}=1.3\%$
where $P_{g/b}$ is the probability to misidentify a gluon as a $b$-jet for a 
$b$-tagging efficiency of 60\%.

%Opakujeme, ale je to potreba kvuli prip. dotazum referee
As discussed in detail in \cite{diffH}, for background calculations, we use 
only an approximate formula. A more realistic approach is to implement
all the background processes in a Monte Carlo program and to perform an
analysis at the detector level (as it was done for the signal
process). 
However none of the background processes mentioned in \cite{diffH} has been 
implemented in any Monte Carlo event generator so far.

At luminosities greater than a few 10$^{33}$~cm$^{-2}$s$^{-1}$, high-energy 
interactions will be accompanied by a non-negligible amount of soft 
interactions, so called pile-up events. The most dangerous combination arises 
from an overlap of the Non-diffractive dijet event with two additional Single 
diffraction events each having a leading proton inside the acceptance of 
forward detectors. The overlap of these three events can resemble a signal 
event and is the most prominent source of background in the $b\bar b$ channel 
at high luminosity, see \cite{CMS-Totem,FP420,diffH,CLP,ATLASnote} for details.
As established in \cite{CLP,ATLASnote,kp2,kp3} we can expect that this overlap 
background can be brought under control by using dedicated fast-timing proton 
detectors with a few pico-second resolution (see \cite{FP420})
and additional experimental cuts based on the exclusivity of the event.

We emphasize that the signal selection and background rejection cuts as used 
for this analysis (more details about these cuts are given in the CMS-Totem 
document \cite{CMS-Totem}, while trigger strategies and optimal mass windows are
explained in \cite{diffH})
were chosen such that they lead to an optimal compromise between the signal 
yield and background reduction. It is important to mention that the signal and 
background yields from that CMS study are in a fair agreement with those 
reported in two ATLAS analyses \cite{CLP,ATLASnote} which both operated at one 
mass point, $\Mh = 120$~GeV. As explained in \cite{diffH}, the overlap 
background cross section is huge and therefore the background rejection cuts 
are very stringent which leads also to a significant reduction of the signal. 
The remaining background is not zero but similarly to \cite{diffH} and 
\cite{diffH2}, we anticipate improvements in the experimental procedure, for 
example smaller misidentification probability $P_{g/b}$ (cf. the improvement
in the light-quark-b misidentification, $P_{q/b}$, techniques in ATLAS 
\cite{ATLAS-bjet} and in CMS \cite{CMS-bjet}), multivariate 
techniques or better resolution in timing detectors which can reduce 
the overlap backgrounds down to a negligible level. Hence the pile-up effects 
are assumed to be negligible after applying all the cuts suppressing the 
pile-up. 
%Consequently, the remaining pile-up backgrounds are not included in 
%our numerical studies performed in the present paper.

%%%%%%%%%%%%%%%%%%%%%%%%%%%%%%%%%%%%%%%%%%%%%%%%%%%%%%%%%%%%%%%%%%%%%%%%%%%%%%%

\subsection*{New benchmark scenarios}\label{scenarios}

Due to the large number of MSSM parameters, a number of benchmark 
scenarios~\cite{benchmark2,benchmark3} have usually been used for the 
interpretation of MSSM Higgs boson searches at LEP~\cite{LEPHiggsMSSM}, 
Tevatron~\cite{TevHiggsMSSM} and LHC~\cite{LHCHiggsMSSM}, 
such as $\Mhmax$, no-mixing, small $\alpha_{\rm eff}$ and gluo-phobic Higgs 
scenarios. The $\Mhmax$, no-mixing and small $\alpha_{\rm eff}$ scenarios have 
been used to evaluate our first 
results \cite{diffH}. Their updates have been published in \cite{diffH2} 
where also more general benchmark scenarios have been examined, the so called 
CDM scenarios (explained in the previous section).  
%that comply with constraints not only to the Higgs sector of MSSM 
%but also to Electroweak precision observables, B physics observables and 
%abundance of Cold Dark Matter, see for instance \cite{EWPOBPOCDM} and 
%references therein. 
%
%In these scenarios the abundance of the lightest SUSY particle, the lightest 
%neutralino, in the early universe is compatible within the $\MA$--$\tb$ plane
%with the CDM constraints as measured by WMAP. The parameters chosen for the 
%benchmark planes are also in agreement with electroweak precision and 
%$B$-physics constraints, see \citere{CDM} for further details.
%Below we will present the prospects for CED Higgs production also in
%two of these four ``CDM benchmark scenarios'', labeled \pdrei\ and
%\pvier. More details about their definition can be found in
%\citere{CDM}. 

In this paper we investigate prospects for the CEP of Higgs boson in 
the new benchmark scenarios that have been proposed recently in 
\cite{newscenarios}. Here we briefly summarize their main features, while their
detailed description can be found in \cite{newscenarios}. 

\begin{enumerate}
\item \ul{The $\Mhmax$ scenario:}\\[.5em]
Mass of the lightest $\cp$-even Higgs boson is maximised at large $\MA$ for 
a given $\tb$. A slight modification compared to the standard  $\Mhmax$ scenario
is an increase of the gluino mass from 0.8~TeV to 1.5~TeV which follows limits
from direct searches for SUSY particles at LHC \cite{HCP2012}. In this scenario
$\Mh$ is in agreement with the discovery of a Higgs-like state only in a 
relatively narrow strip at rather low $\tb$.
\medskip
\item \ul{The Mhmod+ scenario:}\\[.5em]
The $\Mhmax$ scenario can be modified such that in the decoupling regime the 
$\Mh$ values are close to the observed mass of the Higgs signal over a wide 
region of parameter space. The modification consists of reducing the amount of
mixing in the stop sector, i.e. reducing $|X_t/\msusy|$ ($\msusy$ is the mass
of stop and sbottom) compared to a value 
of $\approx 2$ used in the $\Mhmax$ scenario (Feynman-diagrammatic calculation).
This can be done
for both signs of $X_t$. In this scenario this ratio is reduced to 1.5. 
\medskip 
\item \ul{The Mhmod- scenario:}\\[.5em]
Similarly as for the Mhmod+ scenario but with negative $X_t$, $X_t/M_{\rm SUSY} = 
-1.9$. 
\medskip
\item \ul{The Light stop scenario:}\\[.5em]
This scenario can be regarded as an update of the gluo-phobic scenario used in
the past. Values of $X_t$ in the range $2\msusy < X_t < 
2.5\msusy$ lead to reduced gluon fusion rates. In this scenario $X_t = 2\msusy$
 and $\msusy = 500$~GeV. Such a large value of $|X_t|$ and a 
relatively low value of $\msusy$ necessarily lead to the presence of a 
light stop. 
\medskip
\item \ul{The Light stau scenario:}\\[.5em]
The motivation for this scenario stems from the measured diphoton rate of the 
discovered Higgs boson that is somewhat larger than the expectations for a SM
Higgs. Such an enhanced diphoton rate of the lightest $\cp$-even Higgs boson 
may be achieved in a scenario with light staus if a sufficiently large stau 
mixing, $X_{\tau}$, is present.
\medskip 
\item \ul{The Tau-phobic Higgs scenario:}\\[.5em]
This scenario can be regarded as an update of the small $\alpha_{\rm eff}$ 
scenario 
used in the past. Thanks to propagator-type corrections involving the mixing 
between the two $\cp$-even Higgs bosons of MSSM the Higgs coupling to down-type
fermions can be significantly modified which can approximately be termed via
an effective mixing angle $\alpha_{\rm eff}$. This modification occurs for large
values
of $A_{t,b,\tau}$ parameters (denoting Higgs-stop, Higgs-sbottom coupling and 
soft SUSY-breaking parameter in the scalar tau sector, respectively) and large 
values of $\mu$ (higgsino mass parameter) and $\tb$. 
\medskip
\item \ul{The Low-MH scenario:}\\[.5em] 
In this scenario the observed Higgs boson at ~125.5~GeV is identified with the
heavy Higgs boson of the MSSM and behaves roughly SM-like. In this case the 
Higgs sector is very different from the SM one because all five MSSM Higgs 
bosons would be light. The light $\cp$-even Higgs boson of the MSSM would have 
heavily suppressed couplings to gauge bosons. In this scenario a low value of
$\MA$ is required and therefore there is no point in varying $\MA$. It is thus
fixed to the value of 110~GeV and the $\mu$ is varied instead. We have 
checked that other low values of $\MA$ do not affect the final conclusions. 
The remaining parameters are chosen the same as in the Tau-phobic Higgs 
scenario, with the exception of setting $M_{l3} = 1000$~GeV (denoting soft 
SUSY-breaking parameter in the scalar neutrino sector), while in the 
Tau-phobic Higgs is $M_{l3} = 500$~GeV. Note that very recent ATLAS results
\cite{ATLASchargedH} setting exclusion limits on the mass of light charged 
Higgs boson are not yet considered here. They may result in a further 
enlargement of the existing exclusion regions for this scenario but a 
dedicated analysis of those bounds in this scenario is still necessary to be
carried out. 
\end{enumerate}

We remind that in the scenarios 1--6 the discovered Higgs boson is the 
$\cp$-even lightest Higgs boson and we are looking for its heavier partner, 
while in the scenario 7 the situation is opposite. 

%%%%%%%%%%%%%%%%%%%%%%%%%%%%%%%%%%%%%%%%%%%%%%%%%%%%%%%%%%%%%%%%%%%%%%%%%%%%%%%
\subsection*{Bounds from Higgs searches at LEP and the LHC}

Higgs bosons of the SM and the MSSM have been searched for at
LEP~\cite{LEPHiggsMSSM,LEPHiggsSM}, Tevatron~\cite{TevHiggsSM,TevHiggsMSSM}
and LHC~\cite{JATLAS,JCMS,CouplATLAS,CouplCMS,LHCHiggsMSSM}. In our first 
paper \cite{diffH} results were shown in ($\MA$, $\tb$) planes with regions 
excluded by MSSM Higgs searches at LEP. In the second paper \cite{diffH2} the 
exclusion regions coming from Tevatron MSSM Higgs searches were added. In this 
analysis the Tevatron MSSM exclusion regions are superseded by the LHC MSSM
exclusion regions. They correspond to results from direct searches for MSSM 
Higgs bosons at LHC. The search is pursued mainly via the channels 
($\Phi = h,H, A$):
\begin{itemize}
\item $p p \to \Phi \to \tau^+\tau^-~(\mbox{inclusive})$;
\item $b \bar b \Phi,  \Phi \to \tau^+\tau^- ~~\mbox{or}~~ b \bar b~(\mbox{with}~b\mbox{-tag})$; 
\item $p p \to t \bar t \to H^\pm W^\mp \, b \bar b,~H^{\pm} \to \tau \nu_{\tau}~$;
\item$gb \to H^-t ~~\mbox{or}~~ g \bar b \to H^+ \bar t,~H^\pm \to \tau \nu_\tau.$
\end{itemize}
The ($\MA$/$\MHp$, $\tb$) parameter space is constrained by the absence of any
additional state in these production and decay modes, while the masses of the 
first and second generation scalar quarks and the gluino, and to a lesser 
degree of the stop and sbottom masses are constrained by the absence of 
SUSY particles (see \cite{HCP2012} for a recent summary). 

The bounds obtained at the LEP, Tevatron and LHC (with the exception of the 
ATLAS analysis on the charged Higgs boson \cite{ATLASchargedH} as described 
for the scenario 7) have been implemented into the
Fortran code {\tt HiggsBounds}~\cite{higgsbounds1,higgsbounds2} (linked to 
{\tt FeynHiggs}~\cite{feynhiggs1,feynhiggs2,mhiggslong,mhiggsAEC,mhcMSSMlong} 
to provide the relevant Higgs masses and couplings). The full sets of analyses
used to evaluate the exclusion regions are listed in the manual to the 
{\tt HiggsBounds} program \cite{higgsbounds1,higgsbounds2}.
For any parameter point provided to {\tt HiggsBounds} the code
determines whether it is
excluded at the 95\%~C.L.\ based on the published exclusion data.
These excluded regions from LEP {\em and} the LHC are marked in the
MSSM scenario plots shown below. 
We have used the version {\tt HiggsBounds 4.0.0} for our evaluations.

%%%%%%%%%%%%%%%%%%%%%%%%%%%%%%%%%%%%%%%%%%%%%%%%%%%%%%%%%%%%%%%%%%%%%%%%%%%%%%%

\section{Prospects for neutral $\cp$-even Higgs bosons 
in the new benchmark MSSM scenarios}
\label{sec:bench}

In this section we present the prospects for observing the neutral
$\cp$-even MSSM Higgs bosons in CEP. We display our results
in the ($\MA$, $\tb$) planes for seven benchmark scenarios recently proposed in 
\cite{newscenarios} and briefly specified in the previous section. Also shown 
in the plots are the parameter regions excluded 
by the LEP Higgs searches (as dark shaded (blue) areas) and LHC Higgs-boson 
searches (as lighter shaded (red and eventually pink) areas) as obtained with 
{\tt HiggsBounds}~\cite{higgsbounds1,higgsbounds2}. The SM cross section 
used~\cite{bbhatnnlo} for the normalization within {\tt HiggsBounds} is 
evaluated using the MRST2002 NNLO PDFs~\cite{mrst2002}. The use of the updated 
version, namely MSTW2008~\cite{mstw2008} results in a reduction of the cross 
section by $\sim 20\%$, 
%which translates into weaker bounds on $\tb$ by about $10\%$.
while (as discussed in the previous section) the CTEQ6L PDFs \cite{CTEQ6L} 
which give closest predictions to the results obtained in the exclusive 
$\gamma\gamma$ analysis in CDF \cite{CDFgg} would increase the CEP Higgs cross 
section by a factor of almost 3. 
For each point in the parameter space we have evaluated the relevant Higgs
production cross section times
the Higgs branching ratio corresponding to the decay mode under
investigation. The Higgs-boson masses, the decay branching ratios and
the effective couplings for the production cross sections
have been calculated with the program
\fh~\cite{feynhiggs1,feynhiggs2,mhiggslong,mhiggsAEC,mhcMSSMlong}.
The resulting theoretical cross section has been multiplied by the
experimental efficiencies taking into account detector acceptances,
experimental cuts and triggers as discussed in \cite{diffH}.

Where it is relevant, this procedure has been carried out for two different 
assumptions on the luminosity scenario, see Sect.~\ref{lumiscenarios}, for 
which 
the contours for $3\,\si$ significances have been obtained. In remaining cases
only theoretical production cross sections are evaluated and plotted. In figures
we also plot by the light gray (green) so called region of allowed Higgs 
masses, i.e. $\Mh = 125.5 \pm 3$~GeV for scenarios 1--6 and 
$\MH = 125.5 \pm 3$~GeV for the Low-MH scenario. The total uncertainty of 3~GeV
represents a combination of the experimental uncertainty of the measured mass
value ($\sim 0.6$~GeV) and of the theoretical uncertainty in the MSSM Higgs mass
prediction from unknown higher-order corrections. 

\medskip
In our analysis we concentrate only on the $b \bar b$ decay mode. The reason is
that in the parameter space not excluded by the LHC exclusion regions and combined 
with the region of allowed mass of the discovered Higgs boson (roughly 
$6 < \tb < 10$ and $M_{H/h} > 250$~GeV) the final theoretical cross sections for
the $\tau\tau$ decay mode are lower than those for the $b\bar b$ decay mode. 
%Opak
The CEP process with subsequent decay into
bottom quarks is of particular relevance since this channel may provide a unique
possibility for directly accessing the $hb \bar b$ coupling at LHC%
\footnote{
Another interesting idea to access the $b \bar b$ coupling to the
Higgs boson 
is the production via Higgs-strahlung, $V^* \to VH$ ($V = W^\pm, Z$)
in a strongly boosted system~\cite{bbH-boost,ATLAS-VHbb,CMS-VHbb}.
}%
~although the decay into bottom quarks is by far 
the dominant decay mode
of the lighter $\cp$-even Higgs boson in nearly the whole
parameter space of the MSSM (and it is also the dominant decay of a light 
SM-like Higgs). 
For this reason information on the bottom Yukawa coupling is important for
determining {\em any\/} Higgs-boson coupling at the LHC (rather than
just ratios of couplings), see \cite{HcoupLHCSM,HcoupLHC120,lhc2fc}.

\subsection{Scenarios $\Mhmax$, Mhmod+, Mhmod-, Light stop, Light stau and
Tau-phobic Higgs}

\begin{figure*}[htb!]
\resizebox{0.5\textwidth}{!}{
\includegraphics{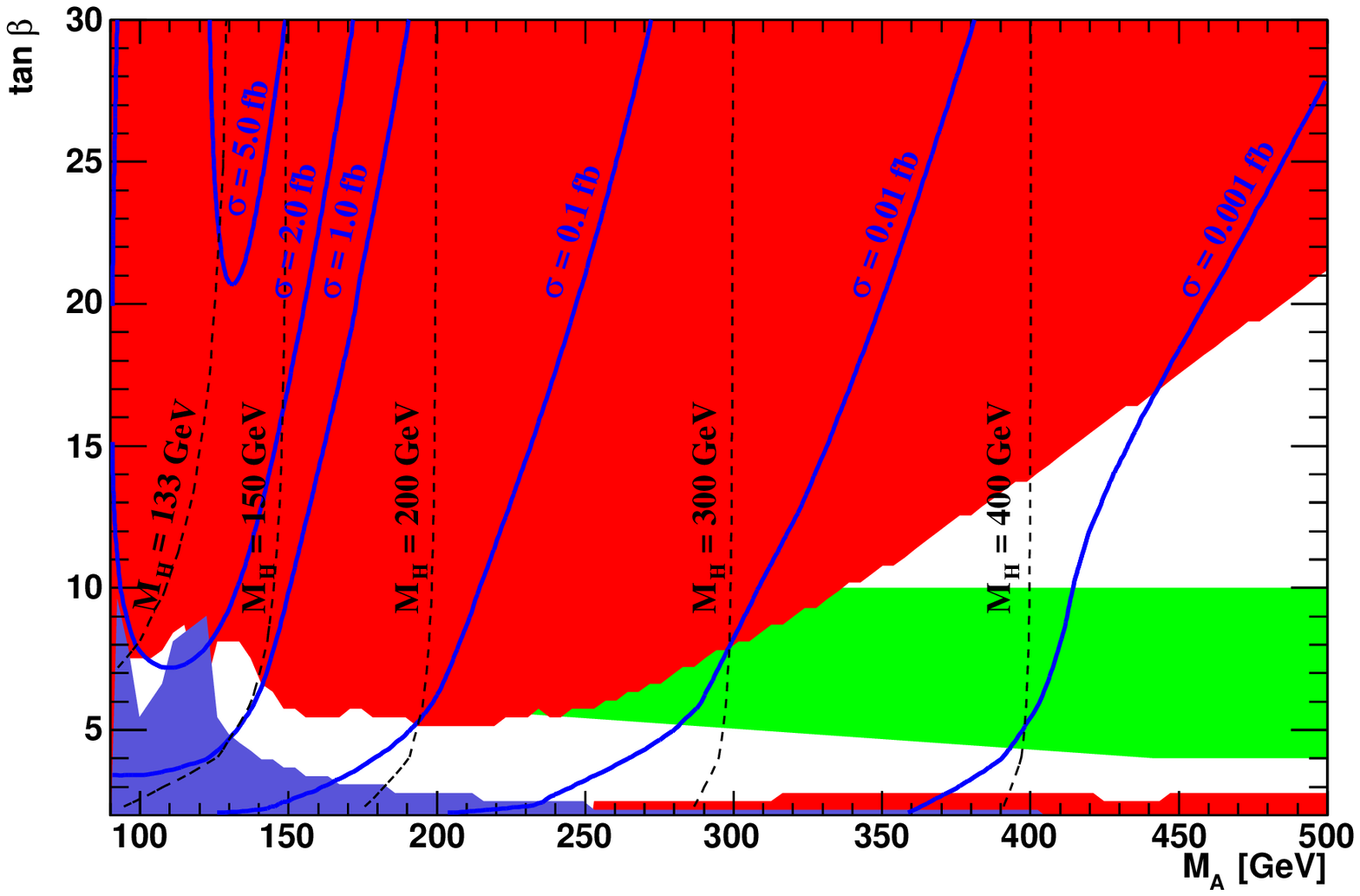}}
\resizebox{0.5\textwidth}{!}{
\includegraphics{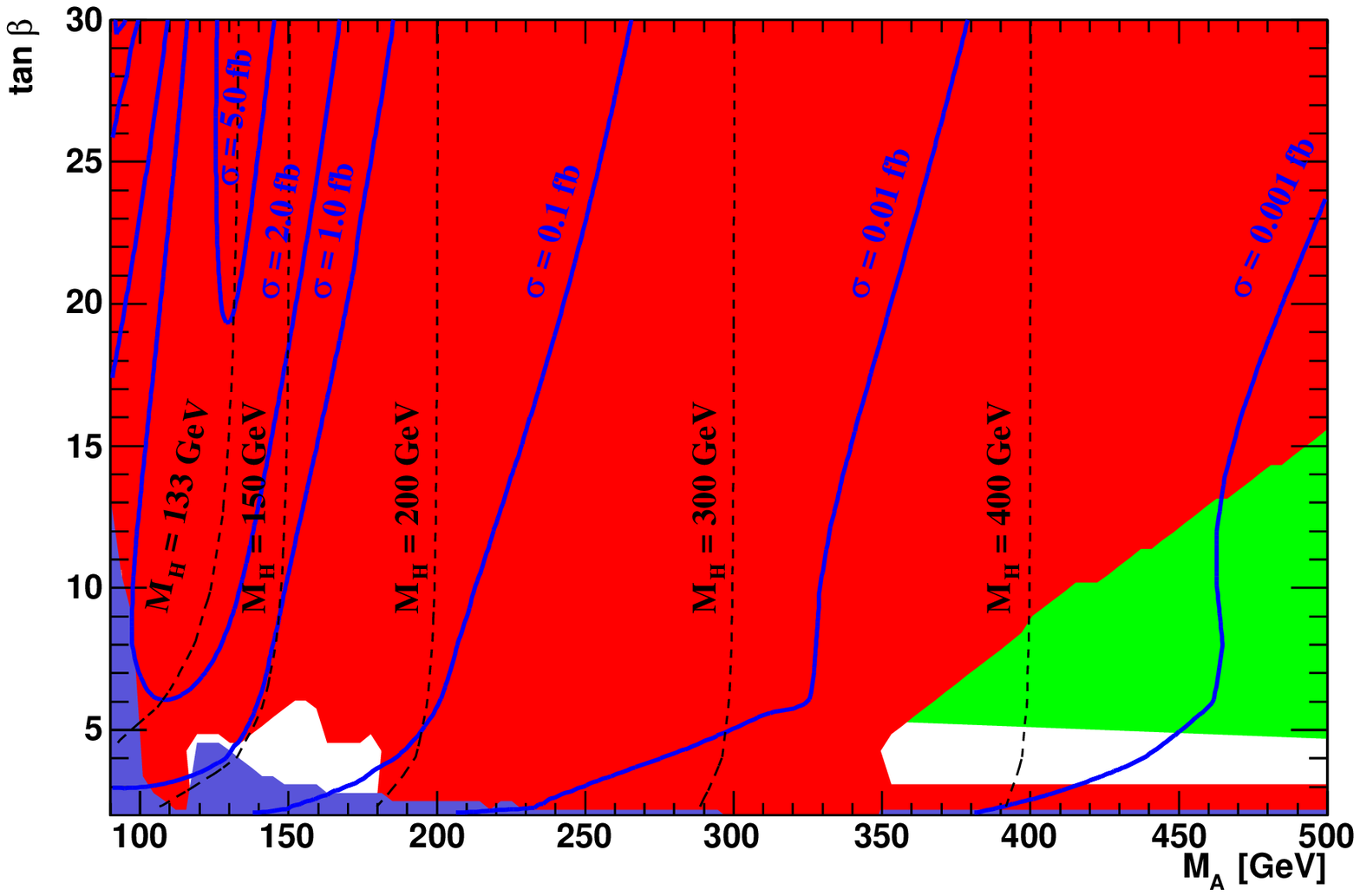}}
\caption{Contours of signal cross sections (solid blue lines) for the 
$H \to b \bar b$ channel in CEP at $\sqrt s = 14$~TeV in 
the ($\MA$, $\tb$) plane of the MSSM within the $\Mhmax$ (left plot) and 
Tau-phobic Higgs (right plot) benchmark scenario. The values of the mass of the 
heavy $\cp$-even Higgs boson, $\MH$, are indicated by dashed (black) contour 
lines. The dark shaded (blue) region corresponds to the parameter region that 
is excluded by the LEP MSSM Higgs searches, the lighter shaded (red) area is 
excluded by the LHC MSSM Higgs searches. The light shaded (green) area 
corresponds to the allowed light Higgs mass region $122.5 < \Mh < 128.5$~GeV. 
}
\label{fig:mhmax}
\end{figure*}

\begin{figure*}[htb!]
\resizebox{0.5\textwidth}{!}{
\includegraphics{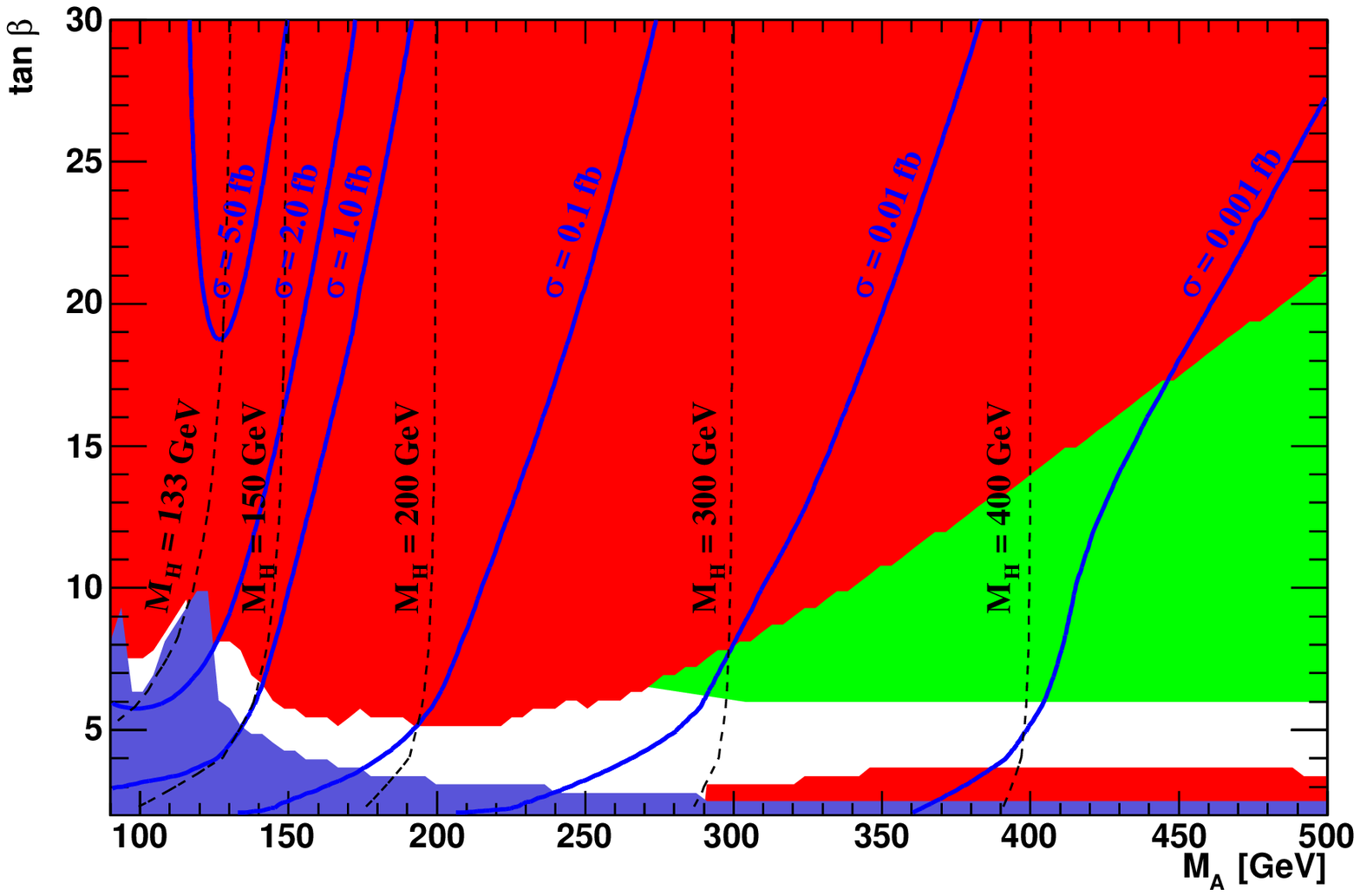}}
\resizebox{0.5\textwidth}{!}{
\includegraphics{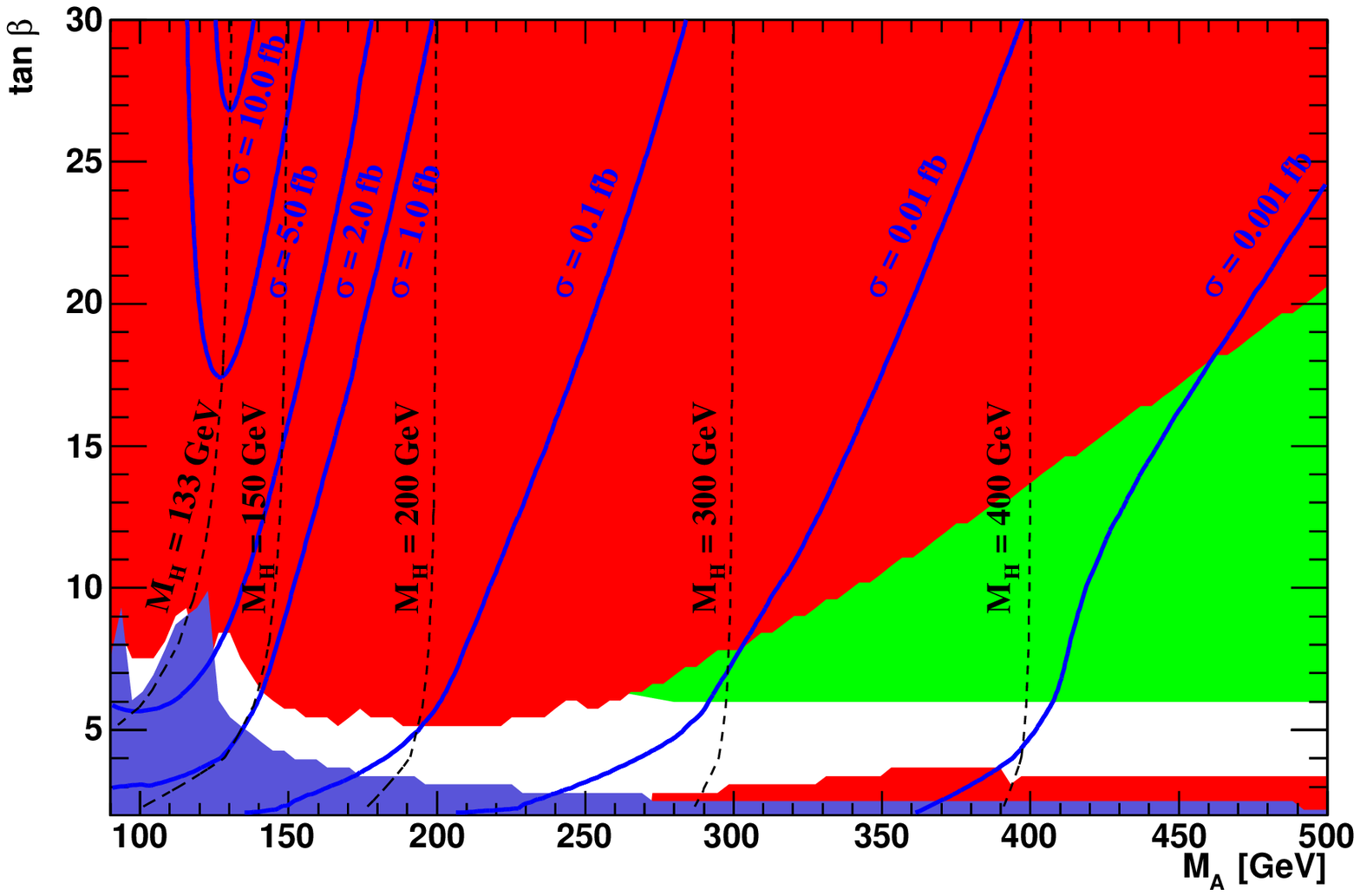}}
\caption{Contours of signal cross sections (solid blue lines) for the 
$H \to b \bar b$ channel in CEP at $\sqrt s = 14$~TeV in 
the ($\MA$, $\tb$) plane of the MSSM within the Mhmod+ (left plot) and 
Mhmod- (right plot) benchmark scenario. The values of the mass of the heavy 
$\cp$-even Higgs boson, $\MH$, are indicated by dashed (black) contour lines. 
The dark shaded (blue) region corresponds to the parameter region that is 
excluded by the LEP MSSM Higgs searches, the lighter shaded (red) area is 
excluded by the LHC MSSM Higgs searches. The light shaded (green) area 
corresponds to the allowed light Higgs mass region $122.5 < \Mh < 128.5$~GeV. 
}
\label{fig:mhmod}
\end{figure*}

\begin{figure*}[htb!]
\resizebox{0.5\textwidth}{!}{
\includegraphics{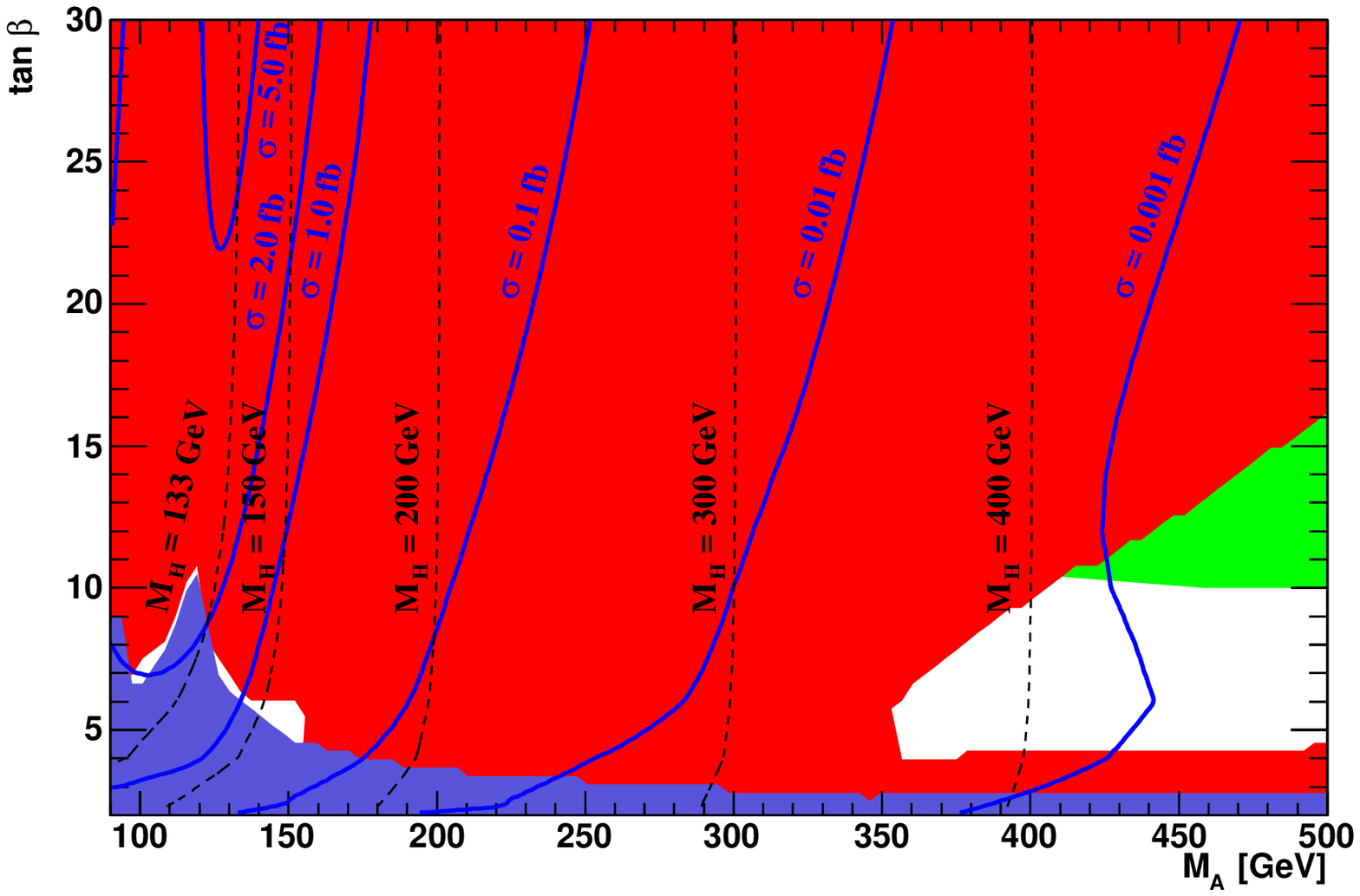}}
\resizebox{0.5\textwidth}{!}{
\includegraphics{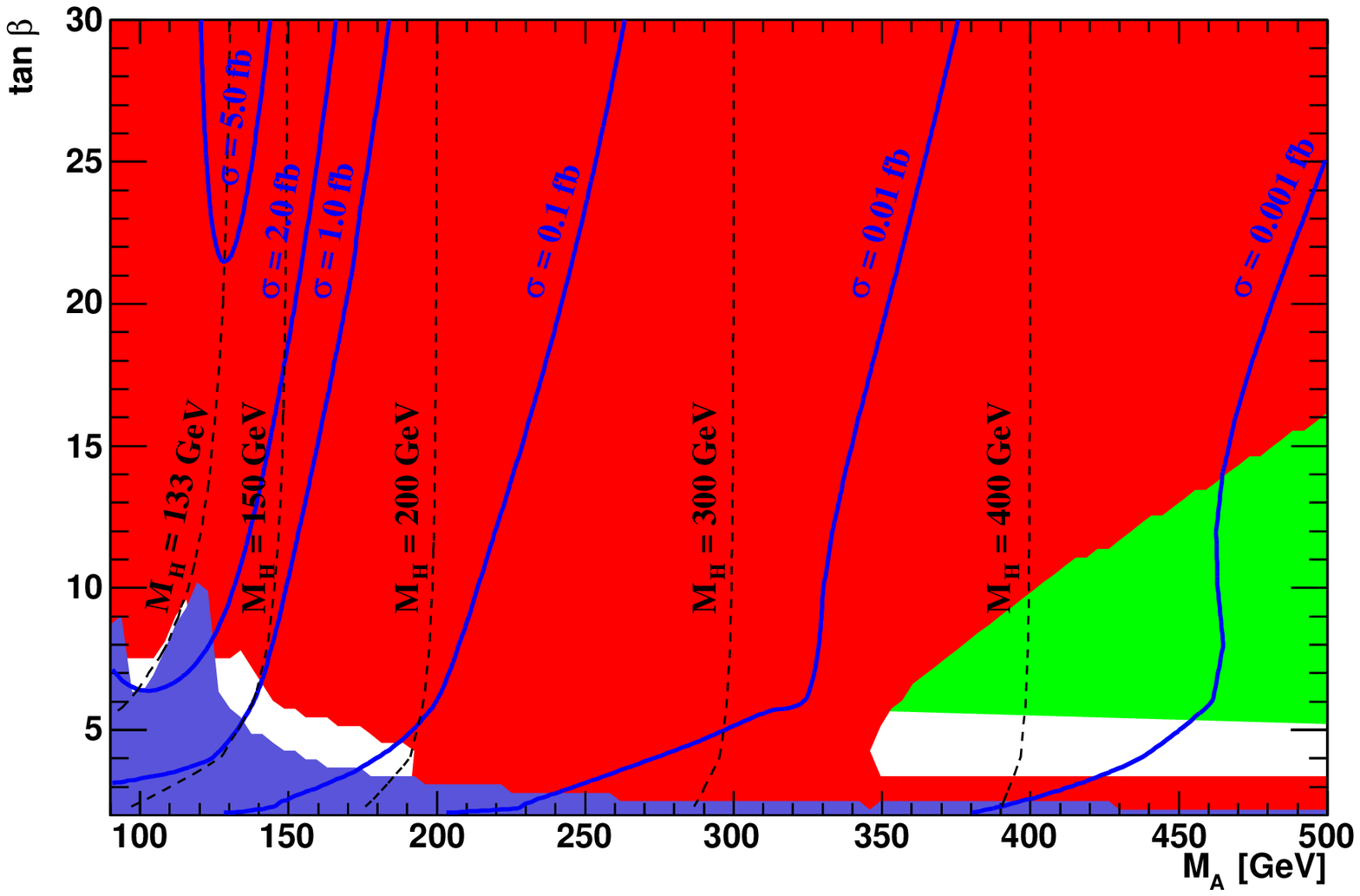}}
\caption{Contours of signal cross sections (solid blue lines) for the 
$H \to b \bar b$ channel in CEP at $\sqrt s = 14$~TeV in 
the ($\MA$, $\tb$) plane of the MSSM within the Light stop (left plot) and 
Light stau (right plot) benchmark scenario. The values of the mass of the heavy 
$\cp$-even Higgs boson, $\MH$, are indicated by dashed (black) contour lines. 
The dark shaded (blue) region corresponds to the parameter region that is 
excluded by the LEP MSSM Higgs searches, the lighter shaded (red) area is 
excluded by the LHC MSSM Higgs searches. The light shaded (green) area 
corresponds to the allowed light Higgs mass region $122.5 < \Mh < 128.5$~GeV. 
}
\label{fig:light}
\end{figure*}

In Figs.~\ref{fig:mhmax}, ~\ref{fig:mhmod} and ~\ref{fig:light} we show the
theoretical cross sections of the $H \to b \bar b$ channel in
CEP in the ($\MA$, $\tb$) plane of the MSSM within the benchmark 
scenarios 1 to 6. We remind that in the scenarios 1--6 the discovered Higgs 
boson is considered to be the lighter one, while we are looking for its heavier
partner. 
We note that the light Higgs behaves more or less SM-like (although the $ggh$ 
coupling can be lower and/or the $hb \bar b$ and $h\tau\tau$ couplings may be 
reduced somewhat), and thus the prospects for the light Higgs are effectively 
those of the SM one (see e.g. \cite{dis12} where we have shown available
cross sections for both the light and heavy MSSM Higgs bosons produced via CEP
and constrained by the Higgs boson discovery and MSSM exclusion regions). 

In Fig.~\ref{fig:mhmax} the results are shown for scenarios $\Mhmax$ (left)
and Tau-phobic Higgs (right), in Fig.~\ref{fig:mhmod} for scenarios Mhmod+ (left)
and Mhmod- (right) and in Fig.~\ref{fig:light} for scenarios Light stop
(left) and Light stau (right). The values of the mass of the heavy $\cp$-even 
Higgs boson, $\MH$, are indicated by dashed (black) contour lines, while the 
values of the theoretical cross sections are indicated by the solid (blue) 
lines. The dark shaded (blue) region corresponds to the parameter region that 
is excluded by the LEP Higgs searches, the lighter shaded (red) areas are 
excluded by LHC Higgs searches. The region of the allowed Higgs mass is plotted 
as a lighter gray (green) band for masses $\Mh = 125.5 \pm 3$~GeV. 

%%%%%%%%%%%%%%%%%%%%%%%%%%%%%%%% Begin FIGURE %%%%%%%%%%%%%%%%%%%%%%%%%%%%%%%%%
%%%%%%%%%%%%%%%%%%%%%%%%%%%%%%%% End FIGURE %%%%%%%%%%%%%%%%%%%%%%%%%%%%%%%%%%%

%%%%%%%%%%%%%%%%%%%%%%%%%%%%%%%% Begin FIGURE %%%%%%%%%%%%%%%%%%%%%%%%%%%%%%%%%
%%%%%%%%%%%%%%%%%%%%%%%%%%%%%%%% End FIGURE %%%%%%%%%%%%%%%%%%%%%%%%%%%%%%%%%%%

%%%%%%%%%%%%%%%%%%%%%%%%%%%%%%%% Begin FIGURE %%%%%%%%%%%%%%%%%%%%%%%%%%%%%%%%%
%%%%%%%%%%%%%%%%%%%%%%%%%%%%%%%% End FIGURE %%%%%%%%%%%%%%%%%%%%%%%%%%%%%%%%%%%

The region of interest is the area of allowed Higgs masses that is not overlaid
by the LHC exclusion region. In different scenarios it covers different phase
space: for the $\Mhmax$ scenario it roughly represents a band delimited by
$\tb$ between 4 and 10 and $\MA$ starting at about 250~GeV. For remaining 
scenarios 2--6 the allowed Higgs mass region represents a wedge which starts at 
($\MA$, $\tb$) = (270~GeV, 6) for the Mhmod+ and Mhmod-, at ($410\,$GeV, 10) for 
the Light stop and at (350~GeV, 5) for the Light stau and Tau-phobic Higgs, and 
reaching $\tb = 16$ at $\MA = 500$~GeV in all scenarios 2--6. In these 
scenarios 
the allowed Higgs mass range covers nearly the whole region not excluded by the 
LHC and LEP MSSM searches and evidently it extends beyond the mass range up to
$\MA = 500$~GeV studied here. We observe that available cross sections in all 
scenarios 1--6 are too small (they are smaller than 0.02~fb) to be considered 
seriously for further studies. 

In general, the CEP of the heavier
$\cp$-even Higgs boson of the MSSM with subsequent decay into bottom
quarks provides a unique opportunity for accessing its bottom Yukawa
coupling in a mass range where for a SM Higgs boson the decay rate into
bottom quarks would be negligibly small. However, due to a massive exclusion 
coming from the LHC MSSM searches, the allowed region concentrates at very low 
$\tb$ values. In this region the enhancement factors MSSM/SM are around 50 (see
\cite{dis12}) but the cross section for the production of the SM Higgs boson 
in CEP is too small. 

%%%%%%%%%%%%%%%%%%%%%%%%%%%%%%%%%%%%%%%%%%%%%%%%%%%%%%%%%%%%%%%%%%%%%%%%%%%%%%%

%%%%%%%%%%%%%%%%%%%%%%%%%%%%%%%% Begin FIGURE %%%%%%%%%%%%%%%%%%%%%%%%%%%%%%%%%
\begin{figure*}[htb!]
\begin{center}
\resizebox{0.8\textwidth}{!}{
\includegraphics{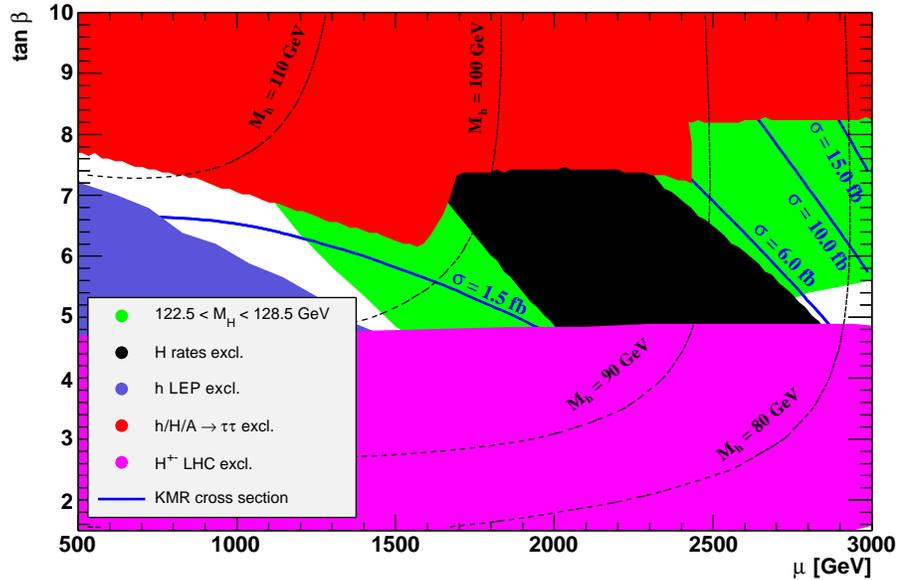}}
\caption{Contours of signal cross section (solid blue lines) for the 
$h \to b \bar b$ channel in CEP at $\sqrt s = 14$~TeV in the 
($\mu$, $\tb$) plane of the 
MSSM within the Low-MH benchmark 
scenario. The values of the mass of the light $\cp$-even Higgs boson, $\Mh$, are
indicated by dashed (black) contour lines. The dark shaded (blue) region 
corresponds to the parameter region that is excluded by the LEP MSSM Higgs 
searches, the lighter shaded (red), the lighter shaded (pink) and black areas 
are excluded by LHC MSSM Higgs searches in the analyses of h/H/A $\rightarrow 
\tau\tau$, charged Higgs and Higgs rates, respectively. The light shaded 
(green) area corresponds to the allowed mass region $122.5 < \MH < 128.5$~GeV. 
}
\label{fig:LowMH1}
\end{center}
\end{figure*}
%%%%%%%%%%%%%%%%%%%%%%%%%%%%%%%% End FIGURE %%%%%%%%%%%%%%%%%%%%%%%%%%%%%%%%%%%

\subsection{Low-MH scenario}
\begin{figure*}[htb!]
\resizebox{0.5\textwidth}{!}{
\includegraphics{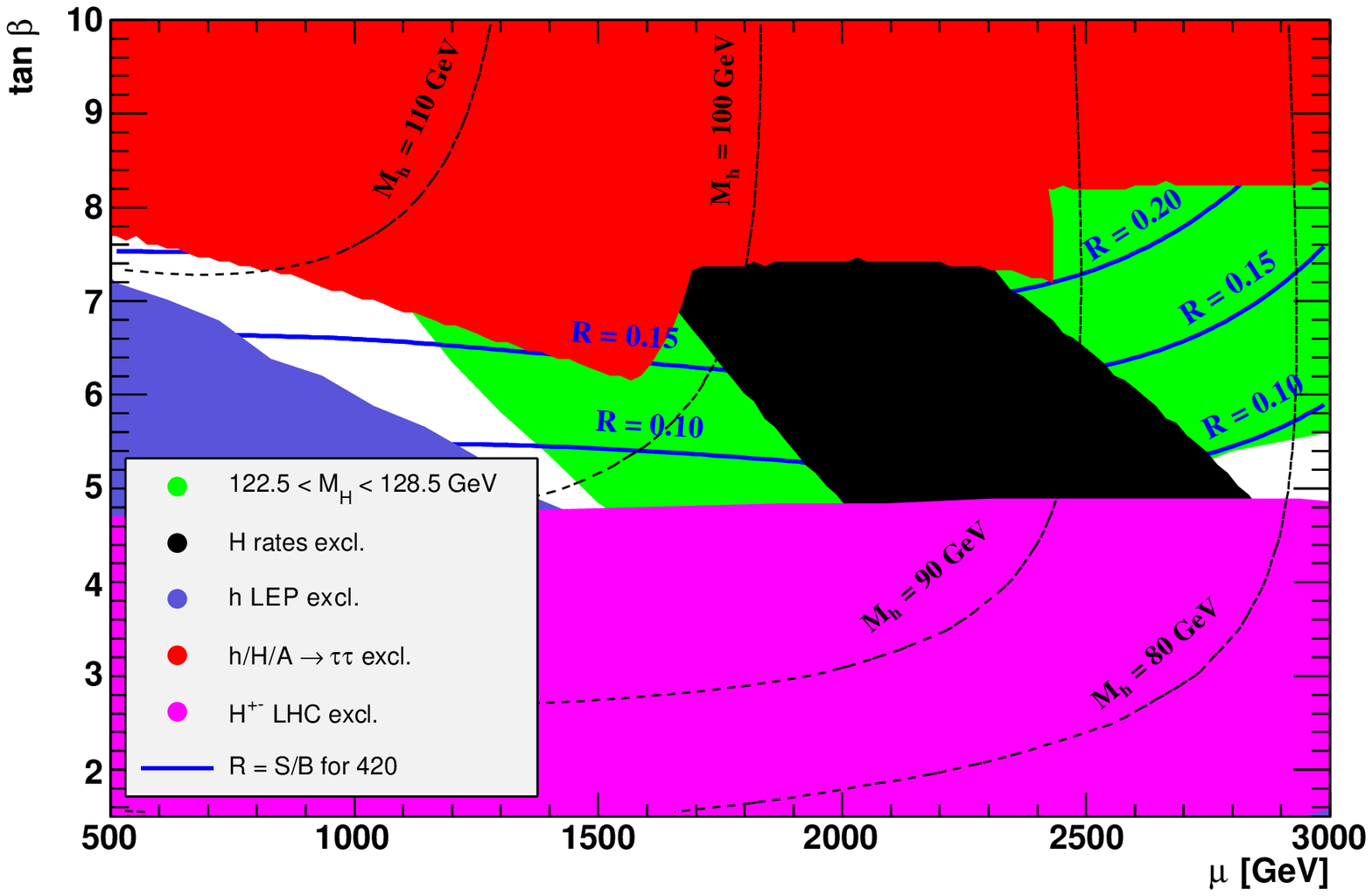}}
\resizebox{0.5\textwidth}{!}{
\includegraphics{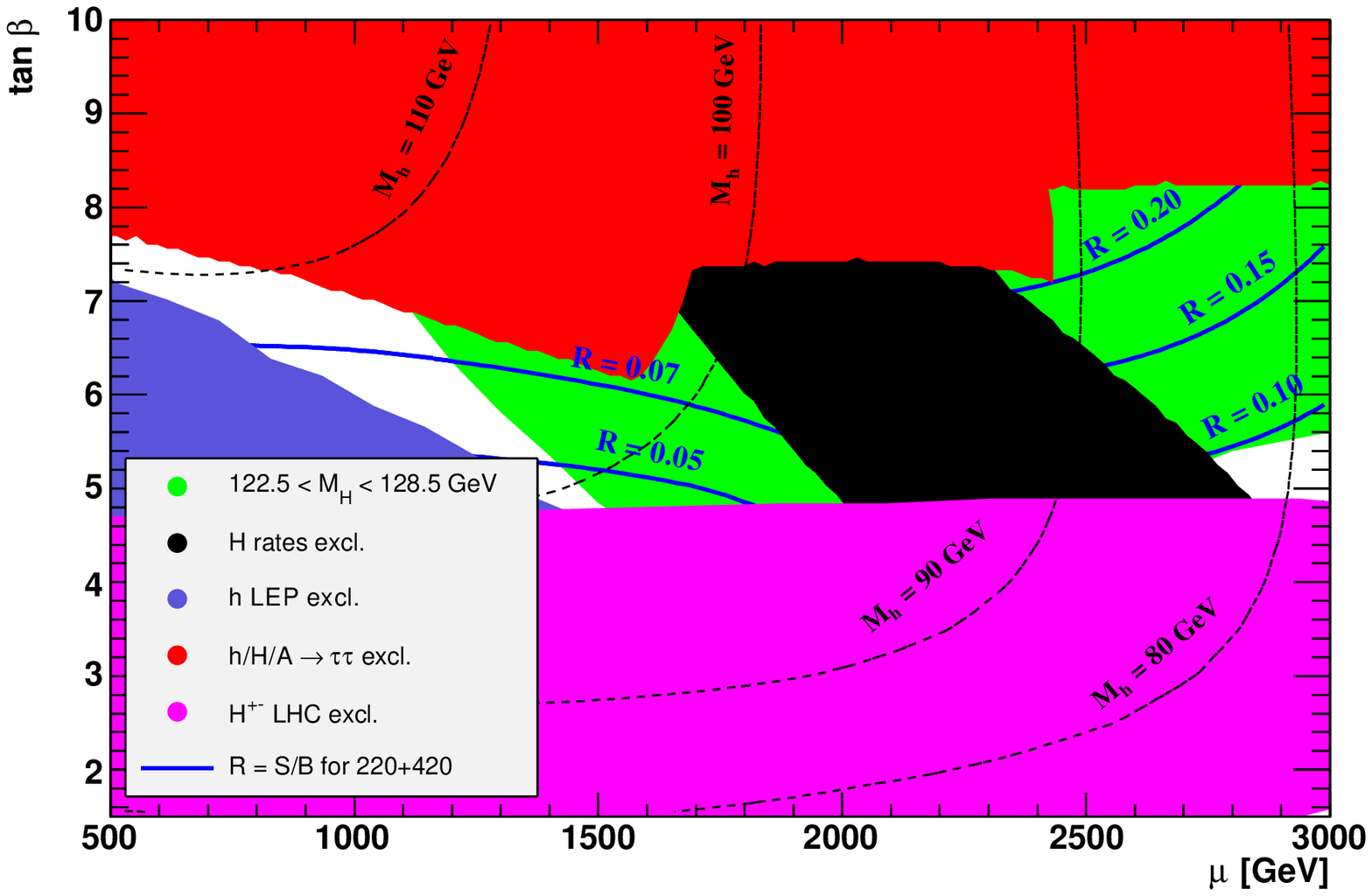}}
\caption{The ratio of signal/background cross sections (solid blue lines) for 
the $h \to b \bar b$ 
channel in CEP at $\sqrt s = 14$~TeV in the ($\mu$, $\tb$) plane of the MSSM within the 
Low-MH benchmark scenario for the 420+420 (Left plot) and the 420+220 (Right 
plot) forward proton detector configuration. In the evaluation of the ratio, 
both signal and background cross sections are multiplied by the total 
experimental efficiencies. The values of the mass of the light $\cp$-even Higgs 
boson, $\Mh$, are indicated by dashed (black) contour lines. The dark shaded 
(blue) region corresponds to the parameter region that is excluded by the LEP 
MSSM Higgs searches, the lighter shaded (red), the lighter shaded (pink) and 
black areas are excluded by LHC MSSM Higgs searches in the analyses of 
h/H/A $\rightarrow \tau\tau$, charged Higgs and Higgs rates, respectively. The 
light shaded (green) area corresponds to the allowed mass region 
$122.5 < \MH < 128.5$~GeV. 
}
\label{fig:LowMH2}
\end{figure*}

\begin{figure*}[htb!]
\begin{center}
\resizebox{0.8\textwidth}{!}{
\includegraphics{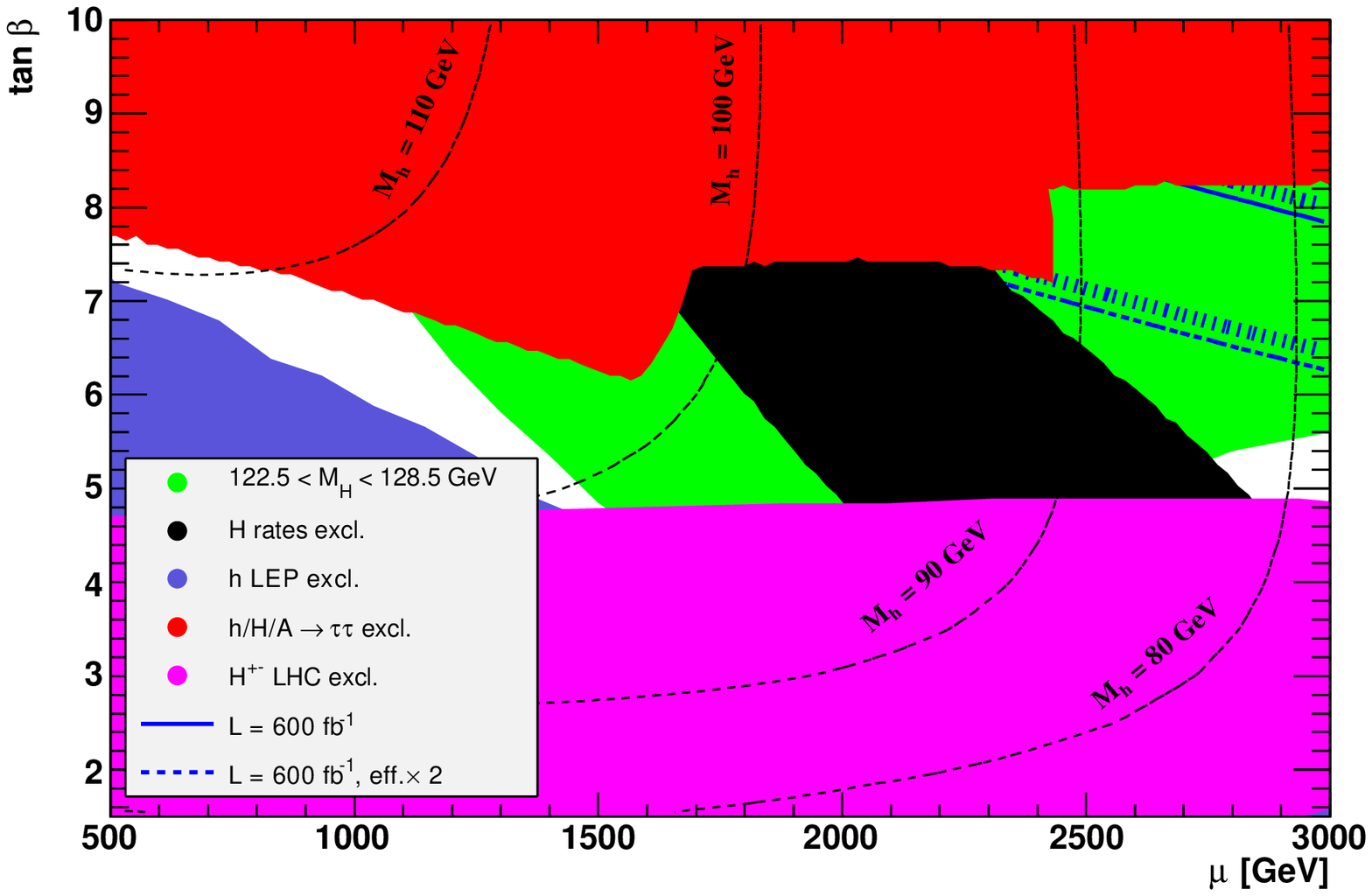}}
\caption{Contours of $3\,\si$ statistical significance (solid blue lines) 
for the $h \to b \bar b$ channel in CEP at $\sqrt s = 14$~TeV in 
the ($\mu$, $\tb$) plane of the MSSM within the Low-MH benchmark scenario. 
The results are shown for assumed effective
luminosities (see text, combining ATLAS and CMS) of \sixooo\ and \sixoooeff.
The values of the mass of the light $\cp$-even Higgs boson, $\Mh$, are
indicated by dashed (black) contour lines. The dark shaded (blue) region 
corresponds to the parameter region that is excluded by the LEP MSSM Higgs 
searches, the lighter shaded (red), the lighter shaded (pink) and black areas 
are excluded by LHC MSSM Higgs searches in the analyses of 
h/H/A $\rightarrow \tau\tau$, charged Higgs and Higgs rates, respectively. The 
light shaded (green) area corresponds to the allowed mass region 
$122.5 < \MH < 128.5$~GeV. 
}
\label{fig:LowMH3}
\end{center}
\end{figure*}

We now turn to the prospects for producing the $\cp$-even Higgs 
boson in CEP channels in the Low-MH scenario. We remind that in this scenario
the discovered Higgs boson is considered to be the heavier one, while we are 
looking for its
lighter partner. As explained in Sect.~\ref{scenarios} the most convenient way 
to present results
is using the ($\mu$, $\tb$) plane rather than the ($\MA$, $\tb$) plane used in 
the scenarios 1--6. In Figs.~\ref{fig:LowMH1}, ~\ref{fig:LowMH2} and 
~\ref{fig:LowMH3} results
for the cross sections, ratios of signal to background cross sections (S/B) and
statistical significances are presented by solid (blue) contour lines. The 
values of the mass of the light $\cp$-even Higgs boson are indicated by dashed 
(black) contour lines. Similarly to the color coding for the
scenarios 1--6, the light shaded (green) band represents the region of allowed
mass of the discovered Higgs boson, in this case of the heavy one, namely 
$122.5 < \MH < 128.5$~GeV. The dark shaded (blue) region represents the LEP MSSM
exclusion region, the lighter shaded (red), the lighter shaded (pink) and black 
areas are excluded by LHC MSSM Higgs searches in the analyses of 
$h/H/A \rightarrow \tau\tau$, charged Higgs and Higgs rates, respectively. As 
described in Sect.~\ref{scenarios}, the results of the ATLAS analysis on the 
charged Higgs bosons \cite{ATLASchargedH} are not included in the exclusion 
regions.  
%%%%%%%%%%%%%%%%%%%%%%%%%%%%%%%% Begin FIGURE %%%%%%%%%%%%%%%%%%%%%%%%%%%%%%%%%
%%%%%%%%%%%%%%%%%%%%%%%%%%%%%%%% End FIGURE %%%%%%%%%%%%%%%%%%%%%%%%%%%%%%%%%%%

%%%%%%%%%%%%%%%%%%%%%%%%%%%%%%%% Begin FIGURE %%%%%%%%%%%%%%%%%%%%%%%%%%%%%%%%%
%%%%%%%%%%%%%%%%%%%%%%%%%%%%%%%% End FIGURE %%%%%%%%%%%%%%%%%%%%%%%%%%%%%%%%%%%

In Fig.~\ref{fig:LowMH1} we show the theoretical signal
cross section for the $h \to b \bar b$ channel in the CEP. 
As can be seen, the signal cross section is an increasing function of $\mu$
reaching its maximum value of about 15~fb at values close to $\mu = 3000$~GeV 
and mass $\Mh$ close to 80~GeV. Note that for normalization purposes at a mass 
of 125~GeV we take for the SM CEP cross section the value of 1.9~fb, see 
\cite{HarlandLang:2013jf} for details. 

In Fig.~\ref{fig:LowMH2} the ratios S/B for two forward proton detector 
configurations, namely the total 420+220 and 420+420 (denoted by 420+220 in 
Fig.~\ref{fig:LowMH2}) and the 420+420 alone, are presented. 
The advantage of plotting this ratio is that various 
uncertainties (such as those connected with the soft survival factor, (N)NLO
effects, choice of proton PDFs etc.) are canceled. To be closer to reality, 
we decided to multiply the theoretical cross sections (those plotted in 
Fig.~\ref{fig:LowMH1}) by the experimental 
efficiencies and integrate over optimal mass windows.
The procedure to get the optimal mass windows from the point of view of the 
highest S/B ratio is described in \cite{diffH}. 

We see that for both forward proton detector configurations, the highest S/B 
values are about 0.2 
and they are located roughly in the same corner as the highest signal cross 
section region seen in Fig.~\ref{fig:LowMH1}. The region of interest is 
therefore ($\mu$, $\tb$) = 
(2500--3000~GeV, 6--8), i.e. ($\mu$, $\Mh$) = (2500--3000~GeV, 80--90~GeV).
In spite of the relatively high signal cross sections, the S/B ratios are 
only moderate. The explanation is twofold: i) theoretical cross sections for 
the irreducible backgrounds are steeply falling functions of mass, see formula 
(8) in \cite{diffH2} composed of mass terms of powers of -6 and -8; 
ii) experimental efficiencies, mass resolution and acceptances are also 
functions of mass, however much less steep and more importantly 
increasing/improving with mass. So in summary, if we move the mass of 
interest from the region of e.g. 120~GeV down to the region of 80--90~GeV, the 
background cross sections greatly increase, while the efficiencies and 
acceptances decrease and the mass resolution slightly worsens (see e.g. 
Fig.~3.7 (right) in \cite{CMS-Totem}). 
In evaluating the S/B ratio, all these experimental effects are properly taken 
into account including the fact that the optimal mass windows for both the
individual 420+420 and 420+220 configurations over which the signal and 
background are 
integrated enlarge with decreasing mass. When comparing these S/B ratios between
the 420+420 and the total sum (i.e. Fig.~\ref{fig:LowMH2} left and right), we 
see that the ratios are identical for masses lower than roughly 100~GeV, while 
they are higher for the 420+420 configuration for masses greater than roughly
100~GeV. This is explained by the mass acceptance of the proposed forward 
proton detectors: in the region $\Mh < 100$~GeV, 
only the 420+420 configuration contributes, while in the region $100 < 
\Mh < 120$~GeV, both configurations are important. 
Due to a much broader optimal mass window for the 420+220 than for the 420+420
in the latter mass range, we pick up much more background than 
with the 420+420 only and hence the contribution of the 420+220 configuration 
makes the total S/B 
%(i.e. the sum of 420+420 and 420+220 contributions to signal and  background) 
smaller than the S/B for the 420+420 configuration alone. 

In Fig.~\ref{fig:LowMH3} we present the contours of $3\,\si$ statistical 
significance for the $h \to b \bar b$ channel in CEP for the 420+220
configuration. The results are shown for assumed effective luminosities of 
\sixooo\ and \sixoooeff. 
%the two lower effective luminosities, namely \sixoo\ 
%and \sixooeff, have given significances smaller than $3\,\si$. 
It is not surprising that the highest achievable
significances are located in the same corner of the green band as have been 
located the highest S/B ratios in Fig.~\ref{fig:LowMH2} (right) plot. We also 
note that this $3\,\si$ statistical significance plot for the 420+420 
configuration is identical to the Fig.~\ref{fig:LowMH3} and therefore it is not
shown. The reason why it is identical is straightforward: the region of 
interest (i.e. significances larger than $3\,\si$) is in the area where Higgs 
masses are very low, namely $80 < \Mh < 90$~GeV. This mass region is unreachable
with the 420+220 configuration, while the mass acceptance of the 420+420 there 
is above 40\% (see e.g. Fig.~3.7 (left) in \cite{CMS-Totem}). We may conclude 
that if the MSSM should be realized as in the Low-MH scenario, i.e. the heavy 
Higgs boson to be seen at mass of 125.5~GeV and the lighter one in the range of 
80--90~GeV, then forward proton detector projects could be very helpful to both 
ATLAS and CMS. At the moment, there are no direct searches at LHC for a Higgs 
boson in such a low-mass region. We emphasize that this 
light Higgs boson could be only seen with stations at 420~m, the less distant 
stations do not contribute.
 
\medskip
A few notes about experimental issues, however, are necessary to make at this 
place. First, from the Fig.~\ref{fig:LowMH3} we see that the total integrated
luminosity needed to observe the light Higgs boson produced in CEP with mass
around 80--90~GeV is of the order of 1000~fb$^{-1}$. Current estimates of
the total amount of acquired data for proposed forward proton detector projects 
at LHC, the AFP and PPS, for realistic data taking scenarios at 
instantaneous luminosities in the range of 2--10 $\times 10^{32}$cm$^2$s$^{-1}$ 
are of the order of 
500~fb$^{-1}$. This means that to observe the light Higgs boson with forward 
proton detectors at this mass range, data from both the AFP and PPS would
have to be combined. Furthermore, putting the AFP stations 
at 420~m into the L1 trigger in ATLAS trigger scheme would be hardly possible 
due to a shorter L1 latency than the time needed for a proton signal in AFP
to be processed and sent back to the Central Trigger Processor. The ways to 
tackle this problem in ATLAS are for example to use existing jet and muon 
triggers at L1 and reducing their rates at L2 by incorporating into L2 the 
track and time information about protons in both 420 stations \cite{CLP} or by 
adding an information about jet $p_T$, $\eta$ and $\phi$ into the L1 
calorimeter \cite{Hbb-triggers}. In CMS the 
current latency is larger than the one of ATLAS and there are plans to even
enlarge it after the Long Shutdown 1. And finally, as was already mentioned, the
total mass acceptance decreases and the mass resolution worsens with decreasing
mass. Another quantity that also deteriorates with decreasing mass is the
b-tagging efficiency. Interestingly the light-quark-b misidentification, 
$P_{q/b}$, improves with decreasing jet transverse energy 
\cite{ATLAS-bjet,CMS-bjet}. 
On the other hand there are areas where a further reduction of the background 
may be expected. We remind that we use $P_{g/b} = $~1.3\% which is at the moment
very conservative given the development seen in ATLAS \cite{ATLAS-bjet} and 
CMS \cite{CMS-bjet} in this area.
Another area with a fair development is fast timing detectors where a sub-10~ps 
resolution in a near future is not excluded, if supported by a proper R\&D 
program. We conclude that investigating 
the mass range of 80--90~GeV with forward proton detectors at LHC is 
more challenging than that around 120~GeV.

%%%%%%%%%%%%%%%%%%%%%%%%%%%%%%%%%%%%%%%%%%%%%%%%%%%%%%%%%%%%%%%%%%%%%%%%%%%%%%%
%%%%%%%%%%%%%%%%%%%%%%%%%%%%%%%%%%%%%%%%%%%%%%%%%%%%%%%%%%%%%%%%%%%%%%%%%%%%%%%

\section{Conclusions}
\label{sec:conclusions}

We have re-analysed in this paper the prospects for probing the Higgs
sector of MSSM with central exclusive Higgs-boson production
processes at the LHC, utilizing forward proton detectors proposed to be 
installed at 220~m and 420~m distance around ATLAS and/or CMS.
The analysis has been performed for CEP of the neutral $\cp$-even 
Higgs bosons $h$ and $H$ of the MSSM and their decays into bottom quarks.

Changes with respect to previous papers \cite{diffH,diffH2} are the 
following. Firstly the Tevatron MSSM exclusion regions shown in \cite{diffH2} 
have been superseded by the LHC MSSM exclusion regions extracted from the most
recent compilation of the LHC MSSM Higgs boson searches.
Secondly the LHC Higgs boson discovery announced last year is 
accounted for in figures by putting the allowed Higgs boson mass band 
$122.5 < M_{h/H} < 128.5$~GeV. And thirdly the results are presented in seven 
new recently proposed low-energy MSSM benchmark scenarios which all are 
compatible with the mass and production rates of the observed Higgs boson signal
at 125.5~GeV. In six scenarios, namely $\Mhmax$, Mhmod+, Mhmod-, Light 
stop, Light stau and Tau-phobic Higgs, the discovered Higgs boson is considered
to be the light $\cp$-even Higgs boson and in the usual ($\MA$, $\tb$) planes 
we investigate the predicted CEP signal of its heavier 
MSSM partner in the allowed Higgs mass band. In the seventh scenario, namely 
the Low-MH, the discovered Higgs boson is considered to be the heavy $\cp$-even 
Higgs boson and
we study the signal of its lighter MSSM partner in the allowed Higgs
boson mass band. Results in this scenario are more convenient to present in the
($\mu$, $\tb$) plane. Compared to last published results in \cite{diffH2}, 
the LEP and LHC exclusion regions rule out most of the MSSM parameter space. The
unexcluded (but at the same time allowed by the LHC Higgs boson discovery) 
region concentrates into a wedge area of small $\tb$, $\tb < 15$ at 
$\MH=500$~GeV where the $\tb$ threshold decreases with decreasing value of 
$\MH$. 

We find that in scenarios 1--6 the theoretical CEP signal cross sections are
too small to produce a detectable signal within a reasonable time scale for 
making use of forward proton detectors. 

%We however stress that the $\Mhmax$, 
%Mhmod+ and Mhmod- are scenarios with an open access to the $b \bar b$ (and 
%$\tau\tau$) decay mode of exclusively produced Higgs boson hence enabling to
%measure the $Hb \bar b$ (and $H\tau\tau$) couplings via detecting the SM-like
%light Higgs boson. 

%There is certainly room
%for improvements in the experimental procedure, let us mention e.g. an expected
%improvement of the gluon-b misidentification probability $P_{g/b}$ compared to 
%the 1.3\% used in our original paper, a sub-10~ps resolution in timing 
%detectors or the use of multivariate techniques. 

The Low-MH scenario gives a much more promising sensitivity. We observe that 
the highest cross sections, S/B ratios and statistical significances are 
located in the corner ($\mu$, $\tb$) = (2500--3000~GeV, 6--8), i.e. 
($\mu$, $\Mh$) = (2500--3000~GeV, 80--90~GeV) which is still not excluded by 
LHC MSSM Higgs searches (as mentioned before, the recent ATLAS results on
the charged Higgs boson \cite{ATLASchargedH} are not included) and lies
in the allowed mass range set up by the Higgs boson discovery at 125.5~GeV.
We find that a light Higgs boson of mass 80--90~GeV decaying into $b \bar b$
and produced via CEP can be observed with $3\,\si$ significance if an integrated
luminosity of around 1000~fb$^{-1}$ is provided. Collecting such an amount of 
data would most likely require to combine data from both the ATLAS and CMS 
forward proton detectors. The low mass state of 80--90~GeV can only 
be detected with forward proton detectors placed at 420~m from the interaction
point, less distant stations do not
contribute due to zero mass acceptance in this mass region. We also comment on 
the challenge to detect reliably a Higgs boson candidate at such a low mass. 
Besides the total mass acceptance decreasing and the mass resolution worsening
also the b-tagging efficiency deteriorate as mass decreases. On the other hand,
we believe that there is still room for improvement of experimental techniques,
let us mention e.g. the expected improvement of the gluon-b misidentification 
probability $P_{g/b}$ compared to the 1.3\% used in our original paper, a 
sub-10~ps resolution in timing detectors or the use of multivariate techniques. 

It may turn out that the observed Higgs boson at mass of 125.5~GeV is of the 
purely SM nature. Then for the CEP and forward proton detector projects to be 
useful to ATLAS and CMS in measuring properties of the SM Higgs boson, the S/B 
ratio needs to be considerably increased. All improvements mentioned above are 
also applicable to the mass region around 120~GeV and will certainly lead to a 
non-negligible increase of S/B.   

A striking feature of CEP Higgs-boson remains that this channel 
provides a valuable information on the spin and the coupling structure of 
Higgs candidates at the LHC. We emphasize that the $J_z = 0$, $\cC$-even, 
$\cP$-even selection rule of the CEP process enables us to estimate 
very precisely (and event-by-event) the quantum numbers of any resonance 
produced via CEP. 

Finally we remind that the proposed forward proton detectors would offer an
unique possibility to probe the $\cp$-violation in the Higgs sector by 
measuring the triple-product correlations of outgoing and incoming proton 
momenta \cite{HarlandLang:2013jf,kmrcp}. 
%%%%%%%%%%%%%%%%%%%%%%%%%%%%%%%%%%%%%%%%%%%%%%%%%%%%%%%%%%%%%%%%%%%%
%%%%%%%%%%%%%%%%%%%%%%%%%%%%%%%%%%%%%%%%%%%%%%%%%%%%%%%%%%%%%%%%%%%%

\subsection*{Acknowledgments}
The work was supported by the project LG13009 of the Ministry of Education of 
the Czech republic.
The author wishes to thank Sven Heinemeyer and Valery Khoze who participated
at the early stage of this analysis for their encouragement and assistance.

%%%%%%%%%%%%%%%%%%%%%%%%%%%%%%%%%%%%%%%%%%%%%%%%%%%%%%%%%%%%%%%%%%%%%%%%%%%%%%%
%%%%%%%%%%%%%%%%%%%%%%%%%%%%%%%%%%%%%%%%%%%%%%%%%%%%%%%%%%%%%%%%%%%%%%%%%%%%%%%

%\newpage

\end{document}

% end of file template.tex